\documentclass[prl,twocolumn,floatfix]{revtex4}
\usepackage{amssymb}
\usepackage{amsmath}
\usepackage{epsfig}
\usepackage{graphicx}

\begin{document}

\title{ Fractional Aharonov-Bohm oscillation of a two-layer ring with two
electrons}

\author{Y.Z. He and C.G. Bao\footnote{The corresponding author}}

\affiliation{State Key Laboratory of Optoelectronic Materials and
Technologies, and Department of Physics, Zhongshan University,
Guangzhou, 510275, P.R. China}

\begin{abstract}
When a circular ring suffers a special topological transformation, it
becomes a two-layer ring. Due to the special topology of the two-layer ring,
orbital angular momenta are allowed to be a half-integer, this would affect
the traditional Aharonov-Bohm oscillation (ABO). In this paper the
fractional ABO of the ground state energy, persistent current, and dipole
transition of a two-layer ring with two electrons has been studied.
Collective and internal coordinates $\theta _{C}$ and $\varphi $ have been
introduced. Based on them a very simple formula for the current has been
obtained, the symmetry constraint imposed on the dipole transition has been
clarified, a strict relation between the photon energies of the dipole
radiation and the persistent current of the ground state has been found.
Comparing with the one-layer rings, the period of the fractional ABO of the
two-layer rings becomes much shorter.
\end{abstract}

\maketitle

\textbf{(1) Introduction}

The quantum ring$^{1}$ is an important member of micro-devices and
is promising in application. Therefore it has been extensively
studied theoretically and experimentally in recent years. It is well
known that the quantum ring has special properties due to the
special ring geometry. \ A distinguished feature is the
Aharonov-Bohm oscillation (ABO) and the fractional ABO (FABO) of the
eigen-energies and persistent currents$^{2-5}$. \ On the other hand,
due to the progress in experimental techniques, distorted rings
(say, a ring undergoes a topological transformation) with specific
geometries containing a given number of electrons can be in
principle fabricated. It is believed that, after a topological
transformation, the physical properties would be accordingly
changed. This is a way to control the properties of micro-devices,
therefore the effect of topological transformations on the rings is
worthy to be studied.

In this paper we shall consider a special case of topological
transformation. Let a one-dimensional ring be twisted into a
"8"-shape as shown in Fig.1a, where upper and lower circles have the
same size. Then, this shape is further bended around the point
$P_{2}$ so that $P_{1}$ and $P_{3}$ are close to each other as shown
in Fig.1b. This system is called a two-layer ring. In such a device,
when an electron in the lower (upper) layer is passing through the
terminal $P_{2},$ it must go to the upper (lower) layer, and this is
the only one choice. \ Although the two layers are close to each
other (the interdistance is considered as zero in the following
calculation), the electrons are not allowed to penetrate from one
layer to the other one except at $P_{2}$. Thus, the domain of the
azimuthal angle $\theta _{i}$ of each electron is no more $[0,2\pi
]$ but $ [0,4\pi ]$. Accordingly, the periodicity of the system has
been changed. \ This change definitely will cause a number of
physical consequences , e.g., orbital angular momenta with
half-integers are allowed as shown below. The aim of this paper is
to clarify the effect of the above topological transformation on the
energy spectra, particle correlations, persistent currents, and
optical properties of the system. \ As a first step, the two-layer
ring is assumed to contain two electrons (the simplest case having
electron-electron correlation). \ Comparison with the results of the
usual one-layer rings is made. The emphasis is to demonstrate how
the (F)ABO is affected.
\begin{figure}[tbph]
\centering
\includegraphics[totalheight=1.4in,trim=30 40 5 10]{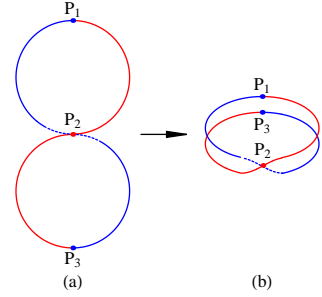}
\caption{(color online). A two-layer ring is transformed from that
(a) a one-dimensional ring is twisted into a "8"-shape where upper
and lower circles have the same size. (b) Then this shape is further
bended around the point $P_{2}$ so that $P_{1}$ and $P_{3}$ are
close to each other. }
\end{figure}

Let a magnetic field $B$ be applied perpendicular to the plane of the rings.
The Hamiltonian of the two-layer ring with two electrons reads
\begin{equation}
\begin{array}{lll}
H=T+V_{12}+H_{Zeeman} &  &  \\
T=\sum_{j=1}^{2}G(-i\frac{\partial }{\partial \theta _{j}}+\Phi )^{2},\ \ \
\ \ G=\frac{\hbar ^{2}}{2m^{\ast }R^{2}} &  &
\end{array}
\end{equation}
where $\theta _{j}$ is the azimuthal angle of the $j$-th electron with a
range from 0 to $4\pi $, $\Phi =\pi R^{2}B/\Phi _{0}$, where $\Phi _{0}=hc/e$
is the flux quantum, $m^{\ast }$ is the effective mess. $H_{Zeeman}$ is the
well known Zeeman energy. $V_{12}$ is the Coulomb interaction which can be
adjusted as $^{6}$
\begin{equation}
\begin{array}{lll}
V_{12}=\frac{e^{2}}{2\epsilon \sqrt{d^{2}+R^{2}\sin ^{2}\frac{\theta
_{1}-\theta _{2}}{2}}} &  &
\end{array}
\end{equation}
where $\epsilon $ is the dielectric constant, and the parameter $d$ is
introduced in ref.[6] to embody the effect of finite width of the ring.

To diagonalize the Hamiltonian, we introduce a set of basis functions
\begin{equation}
\begin{array}{lll}
\phi _{k_{1}k_{2}}=\frac{1}{4\pi }e^{i(k_{1}\theta _{1}+k_{2}\theta _{2})/2}
&  &
\end{array}
\end{equation}
where $k_{1}$ and $k_{2}$ must be integers to assure the periodicity
and $ (k_{1}+k_{2})/2$ is just the total orbital angular momentum
$L$, which is a good quantum number. Due to the special topology as
mentioned, $L$ is allowed to be a half-integer, this is greatly
different from usual one-layer rings. $\phi _{k_{1}k_{2}}$ must be
further anti-symmetrized (symmetrized) when the total spin $S$ is 1
(0).

In this paper $meV$, $nm$ and $Tesla$ are used as units, $m^{\ast
}=0.063m_{e}$, $\epsilon =12.4$ (for InGaAs), $R=30$, $d=0.05R$ is adopted.
The magnitude of $d$ is not sensitive to the qualitative results. The amount
of basis functions should be large enough to assure the accuracy. Actually,
we found that about fifteen hundreds basis functions are enough (to have six
effective figures). The numerical results and related analysis are shown as
follows.

\

\textbf{(2) Separability of the Hamiltonian and the spectra}

We define
\begin{equation}
\begin{array}{lll}
\theta _{c}=\frac{\theta _{1}+\theta _{2}}{2},\ \ \ \ \ \ \varphi =\theta
_{2}-\theta _{1} &  &
\end{array}
\end{equation}
to describe the collective and internal motions, respectively. Then the
Hamiltonian is rewritten as
\begin{equation}
\begin{array}{lll}
H=H_{coll}+H_{int} &  &
\end{array}
\end{equation}
where
\begin{equation}
\begin{array}{lll}
H_{coll}=\frac{1}{2}G(-i\frac{\partial }{\partial \theta _{c}}+2\Phi
)^{2}+H_{Zeeman} &  &  \\
H_{int}=2G(-i\frac{\partial }{\partial \varphi })^{2}+V_{12} &  &
\end{array}
\end{equation}
The collective Hamiltonian $H_{coll}$ depends only on $\theta _{c}$ and is
equivalent to the Hamiltonian of a single particle with a double mass and a
double charge. The internal Hamiltonian $H_{int}$ depends only on $\varphi $
and is irrelevant to $B$. It implies that the eigen-states of $H_{int}$,
namely, the internal states, do not depend on $B$.

The basis functions can be rewritten as
\begin{equation}
\begin{array}{lll}
\phi_{k_1 k_2}=\frac{1}{4\pi} e^{iL\theta_c} e^{i\frac{k_2 - k_1}{4}\varphi}
&  &
\end{array}
\end{equation}
The separability of the Hamiltonian leads to the separability of the
eigen-energies and eigen-states. Thereby the eigen-energy
\begin{equation}
\begin{array}{lll}
E=\frac{1}{2}G(L+2\Phi)^2 + E_{int} - S_z\mu\Phi &  &
\end{array}
\end{equation}
where the first term $\frac{1}{2}G(L+2\Phi )^{2}$ is the kinetic energy of
the collective motion, the second term $E_{int}$ is the internal energy and
the third term $-S_{z}\mu \Phi $ is the well know Zeeman energy. In our
units, $G=604.8/R^{2}$, $\mu =33.53/R^{2}$. The eigen-state with the good
quantum number $L$ reads
\begin{equation}
\begin{array}{lll}
\Psi=\frac{1}{\sqrt{4\pi}} e^{iL\theta_c} \psi_{int} &  &
\end{array}
\end{equation}
where $\frac{1}{\sqrt{4\pi }}e^{iL\theta _{c}}$ describes the collective
motion, while $\psi _{int}$ is the normalized internal state depending only
on $\varphi $. The eigen-states having the same $L$ and $S$ may be different
in $\psi _{int}$, they form a series. From now on the label $(L,S,i)$
denotes the $i$-th state of the series. In particular, the $i=1$ state is
called the first-state of the series, they are candidates of the ground
state depending on the flux $\Phi $. After the diagonalization of the
Hamiltonian, the numerical results are obtained as follows
\begin{figure}[htbp]
\centering
\includegraphics[totalheight=3.5in,trim=30 40 5 10]{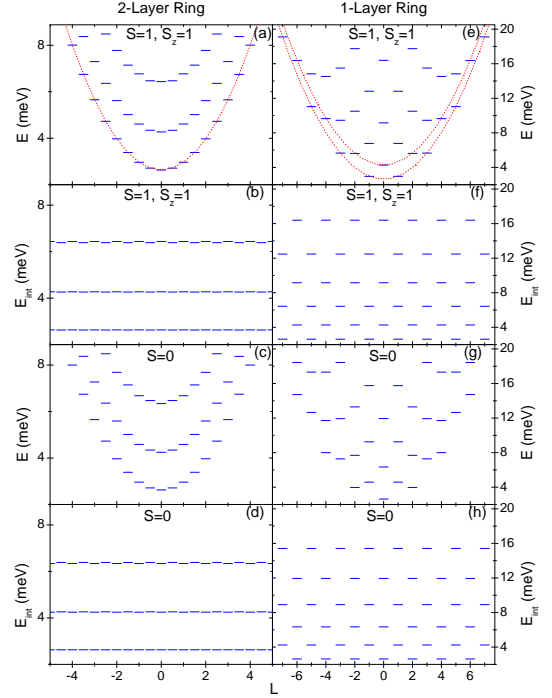}
\caption{(color online). The spectra of a two-layer ring (left
column) and a one-layer ring (right column) against $L$ with $B=0$.
The associated internal energies $ E_{int}$ are also plotted. In 2a
all the $i=1$ levels are joined by a dotted curve to guide the eyes,
this curve splits into two in 2e.}
\end{figure}

When $B=0$, the low-lying eigen-energies $E^{(L,S,i)}$ against $L$
are plotted in Fig.2a (for $S=1$) and 2c (for $S=0$), where, for
each $(L,S)$ -series, the lowest three levels with $i=1,2$, and $3$
are plotted. It was shown in both 2a and 2c that the levels lie
along three parabolic curves (in 2a one of the parabolic curve is
explicitly plotted via a dotted curve). All the levels of each curve
have the same internal energy $E_{int}$, while their collective
energies are equal to $\frac{1}{2}GL^{2},$ this leads to the
parabolic behavior. The associated lowest three internal energies
are plotted in Fig.2b and 2d. The two sets of internal energies with
$S=1$ and $ 0 $, respectively, are one-to-one extremely close to
each other. It implies that the internal structures depend on $S$
very weakly. Fig.2e to 2h are for a one-layer ring (with also two
electrons and with the same $R$) plotted for a comparison. There is
a great difference between the spectra of the two- and one-layer
rings. Firstly, the spectrum of the two-layer ring is much denser
because orbital angular momenta are allowed to be not only integers
but also half-integers, and because the excitation energies of the
internal states are lower (say, comparing 2b and 2f). Secondly, each
of the three internal energies in 2b and 2d splits into two internal
energies as shown in 2f and 2h, one is for $L$ even states, another
for $L$ odd states. Accordingly, each parabolic curve in 2a and 2c
splits into two in 2e and 2g (e.g., the dotted curve in 2a splits
into two dotted curves plotted in 2e). \qquad

The splitting of the internal energies implies that the internal structures
of one-layer rings depend seriously on $(-1)^{L}$. This fact has a profound
symmetry background. It is noted that the most favorable configuration
arises when the two electrons are far away from each other, namely, $|\theta
_{1}-\theta _{2}|=\pi $, where the $e$-$e$ repulsion is minimized. At this
configuration, a spatial interchange of the electrons is equivalent to a
rotation of the system by $\pi $. The former imposes a factor (-1)$^{S}$ on
the wave function due to the fermionic statistics, while the latter imposes
the well known factor $e^{-i\pi L}=(-1)^{L}$. Thus, the favorable
configuration is allowed only if $(-1)^{S}=(-1)^{L}$. Otherwise, the wave
function would contain a node at the configuration.$^{7}$ Obviously, the
appearance of the node would cause an increase of energy, this is the origin
of the splitting as shown clearly in 2f and 2h. In 2f ($S=1$) the $E_{int}$
with $L$ odd is lower, while in 2h ($S=0$) the $E_{int}$ with $L$ even is
lower. The above mechanism, namely, the equivalence of the interchange and
rotation, does not hold for two-layer rings, thus the associated splitting
does not appear in 2b and 2d.
\begin{figure}[tbph]
\centering
\includegraphics[totalheight=3.4in,trim=30 40 5 10]{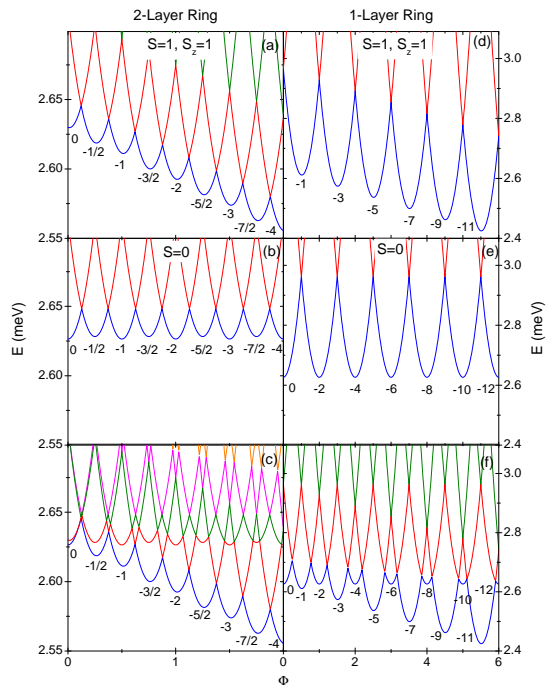}
\caption{(color online). Evolutions of the lowest eigen-energies of
the $S=1$ (a and d) and $ S=0$ (b and e) states against $\Phi $ (the
unit of $\Phi $ is $\Phi _{0}$). c is just the overlap of a and b
for the two-layer ring, while f is the overlap of d and e for the
one-layer ring.}
\end{figure}

When $B\neq 0$, the low-lying eigen-energies against $\Phi $ are
plotted in Fig.3a to 3c to be compared with those of one-layer rings
plotted in 3d to 3f, where the oscillation of the ground state (GS)
energy can be clearly seen. The associated GS angular momentum
$L_{0}$ are marked by the curves. Since the internal energy does not
depend on $B$, when $B$ varies the variation of the GS energy is
caused by the collective term $\frac{1}{2} G(L_{0}+2\Phi )^{2}$,
where the increase of $\Phi $ leads to a decrease of $ L_{0}$ step
by step, this is the mechanism of the (F)ABO. For one-layer rings as
shown in 3f, the decrease of $L_{0}$ is each step by 1, accordingly
the period of $\Phi $ is 1/2. Furthermore, the transition of $L_{0}$
is accompanied by the transition of the GS total spin $S_{0}$ to
keep $ (-1)^{L_{0}}=(-1)^{S_{0}}$. This is a necessary condition for
the GS to be free from the inherent node at the $|\theta _{1}-\theta
_{2}|=\pi $ configuration. However, for two-layer rings, the
decrease of $L_{0}$ is each step by 1/2 as shown in 3c, accordingly
the period of $\Phi $ is 1/4 implying a much denser oscillation.
Besides, the constraint $ (-1)^{L_{0}}=(-1)^{S_{0}}$ is no more
valid, there is no transition of $ S_{0} $, instead, the total spin
remains to be 1 due to the Zeeman energy.

\

\textbf{(3) Structure of internal states}

The density of an internal state is defined as
\begin{equation}
\begin{array}{lll}
\rho _{1}(\varphi )=\psi _{int}^{\ast }(\varphi )\psi _{int}(\varphi ) &  &
\end{array}
\end{equation}

\begin{figure}[htbp]
\centering
\includegraphics[totalheight=2.8in,trim=30 40 5 10]{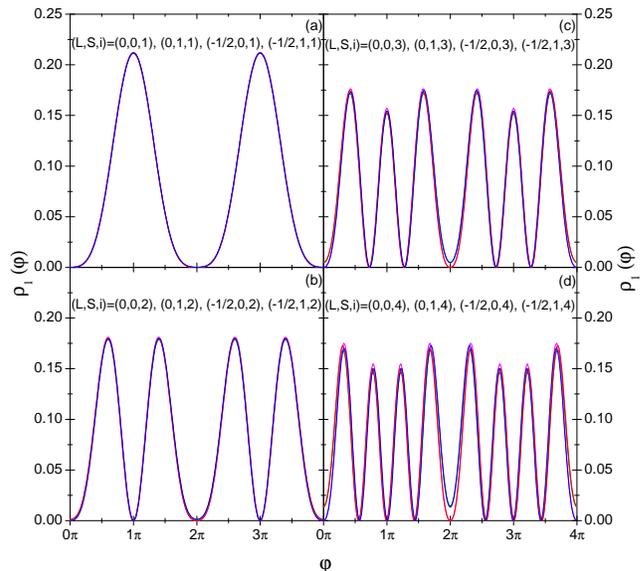}
\caption{(color online). $\protect\rho _{1}$, the density of the
internal states, as a function of $\protect\varphi $.}
\end{figure}

Examples are given in Fig.4 for $i=1$ to 4 states. There are the following
features:

(i) For each case of $i$ the curves with distinct $L$ and $S$ nearly
overlap, it implies that $\psi _{int}(\varphi )$ depends on $L$ and
$S$ very weakly. \ On the contrary, the internal states of one-layer
rings depend on $ (-1)^{L}$ and $S$ very strongly.$^{8}$

(ii) More nodes would be contained if $i$ is larger as clearly shown
from a to d. However, since $\psi _{int}$ must be a superposition of
$e^{i\frac{ k_{2}-k_{1}}{4}\varphi }$ (refer to eq.(7)), therefore
the periodicity of $ \psi _{int}$ depends on $L=(k_{1}+k_{2})/2$.
When $L$ is an integer, the period of $\varphi $ is $4\pi $, when
$L$ is an half-integer, the period of $ \varphi $ is $8\pi $. Thus,
due to the distinction in periodicity, the structures of the
internal states might be different. It was found that the effect of
the periodicity is very weak when $i$ is small. However, when $
i\geq 3$ this effect can be seen clearer as shown in Fig.4c and 4d.
Incidentally, the periodicity would also affect the internal
energies and can be seen if $i$ is larger as shown by the $i=3$
levels in Fig.2b and 2d.

(iii) The favorable configuration, namely, $|\theta _{1}-\theta _{2}|=\pi $
or $3\pi /2$ is possessed by all the first-states distinct in $L$ and $S$.
This is shown in Fig.4a.

\

\textbf{(4) Persistent current}

It is well known that, from the conservation of mass, the current of the
first electron is
\begin{equation}
\begin{array}{lll}
J_{1}=\frac{g}{2}\int [\Psi ^{\ast }(-i\frac{\partial }{\partial \theta _{1}}
+\Phi )\Psi +c.c.]d\theta _{2} &  &
\end{array}
\end{equation}
where $g=\hbar /(m^{\ast }R^{2})$. Since $J_{1}$ does not depend on $\theta
_{2}$, it can be integrated over $\theta _{2}$ and then divided by $4\pi $.
Thus the total persistent current $J=J_{1}+J_{2}$ can be written as
\begin{equation}
\begin{array}{lll}
J=\frac{g}{8\pi }\int [\Psi ^{\ast }(-i\frac{\partial }{\partial \theta _{1}}
-i\frac{\partial }{\partial \theta _{2}}+2\Phi )\Psi +c.c.]d\theta
_{1}d\theta _{2} &  &
\end{array}
\end{equation}
where both $\theta _{1}$ and $\theta _{2}$ are integrated from 0 to $4\pi $.
Since $\frac{\partial }{\partial \theta _{1}}$+$\frac{\partial }{\partial
\theta _{2}}$=$\frac{\partial }{\partial \theta _{c}}$, By using the
arguments $\theta _{c}$ and $\varphi $ and making use of the separability,
the integration can be easily performed, and we have
\begin{equation}
\begin{array}{lll}
J=\frac{1}{4\pi }g(L+2\Phi ) &  &
\end{array}
\end{equation}

\begin{figure}[htbp]
\centering
\includegraphics[totalheight=2.4in,trim=30 40 5 10]{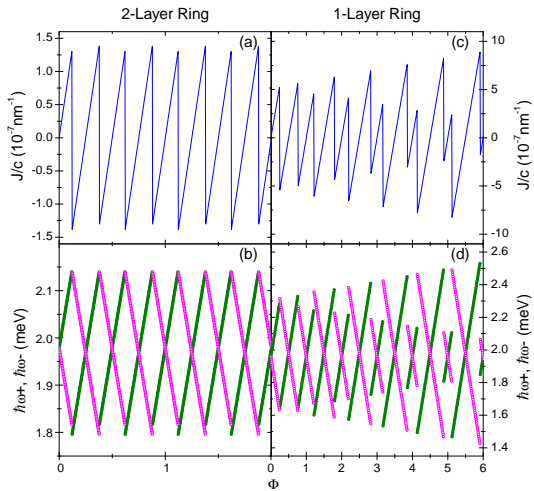}
\caption{(color online). The current $J/c\ (10^{-7}nm^{-1})$ and the
two photon energies of the ground state (a and b) against $\Phi $. c
and d are for the associated one-layer ring just for a comparison.}
\end{figure}

Since the current and the eigen-energy both contain the factor $(L+2\Phi )$,
they have the same mechanism of oscillation (namely, the transition of $L$
each step by 1/2 in accord with $\Phi $) with the same period $\Phi =1/4$,
their oscillations match with each other. The maximum of $J$ arises when $L$
undergoes a transition, we have $|J_{\max }|=g/16\pi $.

On the other hand, for one-layer two-electron rings,
$J=\frac{1}{2\pi } g(L+2\Phi ).$ When $B$ is not large, the period
of $\Phi $ is $1/2$ and the maximal current $|J_{\max }|=g/4\pi $.
When $B$ is sufficiently large and enters into the ABO region (so
that the GS is dominated by $S=1$ states), the period of $\Phi $ is
$1$ and the maximal current $|J_{\max }|=g/2\pi $.$ ^{8}$

Thus, comparing with the one-layer rings, the strength of $J$ of the
two-layer rings is at least four times weaker and the period is at least two
times shorter. This is shown in Fig.5a and 5c. There is one more point
noticeable. For one-layer rings, the FABO will be gradually changed to ABO
during the increase of $B$. This is shown in 5c where the amplitude of the
current increases with $B$ until $B$ arrives at the border of the ABO
region. Such an increase does not appear for two-layer rings.

\

\textbf{\ (5) Dipole transitions and related photon energies}

Let the labels $(init)$ and $(f)$ be used to denote the initial and final
states of a dipole transition. For the transition from $\Psi
_{L,S,i}^{(init)}$ to $\Psi _{L^{\prime },S^{\prime },i^{\prime }}^{(f)}$,
the probability is$^{9}$
\begin{equation}
\begin{array}{lll}
P_{(init),\pm }^{(f)} & =\frac{2e^{2}}{3\hbar }(\frac{\omega }{c})^{3}R^{2}
&  \\
& |<\Psi _{L^{\prime },S^{\prime },i^{\prime }}^{(f)}|(e^{\pm i\theta
_{1}}+e^{\pm i\theta _{2}})|\Psi _{L,S,i}^{(init)}>|^{2} &
\end{array}
\end{equation}
where $\hbar \omega =E_{(f)}-E_{(init)}$ is the photon energy. Since
$e^{\pm i\theta _{1}}+e^{\pm i\theta _{2}}\equiv 2e^{\pm i\theta
_{C}}\cos (\varphi /2)$, during the transition not only the total
orbital angular momentum arising from the collective rotation should
be changed by $\pm 1$, the internal structure would be changed also.
Using the internal degree of freedom and making use of the
separability, the probability can be rewritten as
\begin{equation}
\begin{array}{ll}
P_{(init),\pm }^{(f)}= & \frac{8e^{2}}{3\hbar }(\frac{\omega }{c}
)^{3}R^{2}\delta _{L^{\prime },L\pm 1} \\
& |<\psi _{int}^{(f)}|\cos \frac{\varphi }{2}|\psi
_{int}^{(init)}>|^{2}
\end{array}
\end{equation}

It is noted that, in the domain of $\varphi $ from 0 to $2\pi $,
$\cos \frac{ \varphi }{2}$ is antisymmetric with respect to $\pi $
(i.e., $\cos \frac{\pi -\varphi }{2}=-\cos \frac{\pi +\varphi
}{2}).$ Therefore $\psi _{int}^{(f)\ast }\psi _{int}^{(init)}$ must
also be antisymmetric in this domain, otherwise the associated
integration in the domain would be zero. This statement holds also
for the domain of $\varphi $ from $2\pi $ to $4\pi $, thus the
transition of the internal states might be constrained. It turns out
that, when the initial state $\Psi _{L,S,i}^{(init)}$ and the final
state $\Psi _{L^{\prime },S^{\prime },i^{\prime }}^{(f)}$ together
have $ i+i^{\prime }$ an even integer, $<\psi _{int}^{(f)}|\cos
\frac{\varphi }{2} |\psi _{int}^{(init)}>$ is zero. When
$i+i^{\prime }$ is odd and $|i^{\prime }-i|\geq 3,$ the integration
is not zero but very small due to the nodal structures of the
internal states. Consequently, the transition is essentially
concentrated into the final state with $L^{\prime }=L\pm 1$, $
S^{\prime }=S$, and $|i^{\prime }-i|=1$, the associated probability
depends essentially on the internal structures and the energy
difference $\hbar \omega $.

Let $(o)$ denote the GS. For the GS transition, $L^{\prime }=L_{(o)}\pm 1,\
S^{\prime }=S,\ i^{\prime }=i+1=2$. An example is given in Fig.6, where the
FABO with the period $1/4$ is caused by the oscillation of $\omega $. The
probability of the GS jumping to a higher state (say, $i^{\prime }\geq 4$)
is very small, e.g., when $i^{\prime }=4$, it is smaller by two order.
\begin{figure}[tbph]
\centering
\includegraphics[totalheight=2.0in,trim=30 40 5 10]{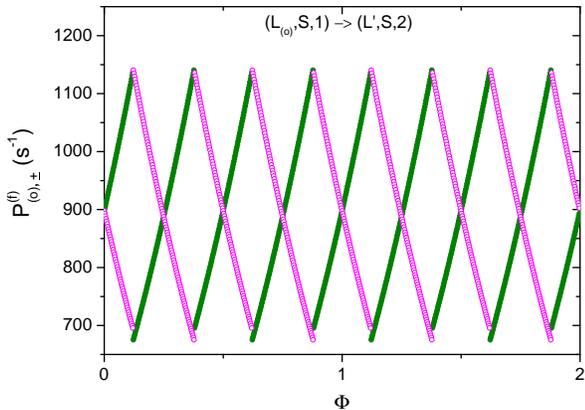}
\caption{(color online). The probability of ground state transitions
from $(L_{(o)},S,1)$ to $(L^{\prime },S,2)$ against $\Phi .$ The
black (white) circles denote the case $L^{\prime }=L_{(o)}+1\
(L_{(o)}-1)$.}
\end{figure}

Let the two photon energies of the main GS transitions ( $i^{\prime }=2$) be
denoted by $\hbar \omega _{+}$ and $\hbar \omega _{-}$ for $L^{\prime
}=L_{(o)}+1$ and $L_{(o)}-1$, respectively. From eq.(8),
\begin{equation}
\begin{array}{lll}
\hbar \omega _{\pm } & =\frac{G}{2}[1\pm 2(L_{(o)}+2\Phi
)]+E_{int}^{(f)}-E_{int}^{(o)} &
\end{array}
\end{equation}
which is plotted against $\Phi $ in Fig.6b. For a comparison, those of a
one-layer ring is plotted in 5d. Both figures have the FABO with periods 1/4
(5b) and 1/2 (5d, when $\Phi $ is small ), respectively. Furthermore, the
maximal amplitude in 5b is a constant, while that in 5d is increasing until
arriving at the border of the ABO region.

We define $\Delta _{\hbar \omega }=\hbar \omega _{+}-\hbar \omega _{-}$. Due
to eq.(13) we have
\begin{equation}
\begin{array}{lll}
\triangle _{\hbar \omega }=2hJ_{(o)} &  &
\end{array}
\end{equation}
where $h$ is the Planck's constant. This relation is a new finding,
it relates $\Delta _{\hbar \omega }$ directly with $J_{(o)}$. The
fact that these two kinds of oscillations match with each other
exactly is a common feature for various narrow rings. For an
example, the one-layer two-electron rings have a roughly similar
relation $\triangle _{\hbar \omega }=hJ_{(o)}$.$ ^{8}$ Since the
photon energy is much easier to be measured accurately, this
relation provides an effective way to measure the currents.

\

\

In summary, we have studied the FABO of the eigen-energies, persistent
current, and dipole transition of the two-layer rings with two electrons.
Collective and internal coordinates $\theta _{C}$ and $\varphi $ have been
introduced. \ Based on them the underlying physics can be better understood.
\ The following points are reminded.\ (i) we have derived a very simple
formula for the current eq.(13), which is different from that arising from
the famous formula $eJ=-\frac{c}{\Phi _{0}}\frac{\partial E}{\partial \Phi }$
for the one-layer ring by a factor 1/2 due to the topology. However, if both
the currents in the two layers are taken into account, their sum recovers
the result of the above famous formula. (ii) the symmetry constraint imposed
on dipole transitions can be revealed (eq.(15), which explains the selection
rule of $i^{\prime }$), (iii) the strict relation between $\triangle _{\hbar
\omega }$ and the current has been established (eq.(17), which is different
from the associated formula derived in [8] for one-layer rings by also a
factor 2), etc.. The two-layer rings do not have the ABO, but only the FABO.
The period of the FABO is found to be two times shorter than that of the
one-layer rings when $B$ is not large, and four times shorter when $B$ has
entered into the ABO region.

The above discussion can be generalized to the case with many layers
(namely, a multi-layer ring). E.g., for a seven-layer ring with two
electrons, fractional orbital angular momenta ($L=I/7$) would emerge, the
period of the FABO becomes $1/14.$ The very short period of the FABO is a
noticeable feature of multi-layer rings.

\

Acknowledgment: The support from the NSFC under the grant 10574163 is
appreciated.

\

REFERENCES

1, S.Viefers, P. Koskinen, P. Singha Deo and M. Manninen, \ Physica E
\textbf{21}, 1 (2004)

2, U.F. Keyser, C. F\"{u}hner, S. Borck, R.J. Haug, M. Bichler, G.
Abstreiter and W. Wegscheider, \ Phys. Rev. Lett. \textbf{90}, 196601 (2003)

3, D. Mailly, C. Chapelier and A. Benoit, \ Phys. Rev. Lett.\textbf{70},
2020 (1993)

4, A. Fuhrer, S. L\"{u}scher, T. Ihn, T. Heinzel, K. Ensslin, W. Wegscheider
and M. Bichler, \ Nature (London) \textbf{413}, 822 (2001)

5, A.E. Hansen, A. Kristensen, S. Pedersen, C.B. Sorensen and P.E. Lindelof,
\ Physica E (Amsterdam) \textbf{12}, 770 (2002)

6, M. Korkusinski, P. Hawrylak and M. Bayer, \ Phys. Stat. Sol. B
\textbf{234 }, 273 (2002)

7, C.G. Bao, W.F. Xie and C.D. Lin, \ J. Phys. B: At. Mol. Opt. Phys.
\textbf{27}, L193 (1994)

8, Y.Z. He and C.G. Bao, \ arXiv:0704.0070v1 [cond-mat.mes-hall]

9, M.E. Rose, \textit{Multipole Fields} (John Wiley and sons, New York, 1955)

\end{document}